# X2
# MOLECULAR ORBITAL THEORY OF THE GASEOUS BOSE-EINSTEIN CONDENSATE:
## *NATURAL ORBITAL ANALYSIS OF STRONGLY CORRELATED GROUND AND EXCITED STATES OF AN ATOMIC CONDENSATE IN A DOUBLE WELL*

William P. Reinhardt and Heidi Perry
*University of Washington, Seattle, WA 98121-1700, U.S.A.*

**Abstract** The possibility, envisaged in 1925 by Einstein following the suggestion of Bose, of a dilute gas of atoms being *condensed* into a single quantum state was experimentally achieved in 1995 following decades of research. An avalanche of experiment and theory has followed, leading to the awarding of the 2001 Nobel Prizes in Physics to three of the pioneering experimentalists. Theory, mostly couched in the language and formalism of condensed matter physics, has developed apace. What we point out here is that a condensate in a double well trap may be thought of exactly as a diatomic molecule with delocalized molecular orbitals (MOs) filled with Bosons, rather than electrons. As in quantum chemistry, configuration interaction must be included to supplement the MO picture if dissociation is to be correctly described. Such dissociation is called *condensate fragmentation* in the condensed matter theory literature. Natural orbitals and their occupation numbers provide a clear and succinct way of determining "where the atoms are" in highly correlated states as fragmentation occurs, and are particularly helpful in understanding what is actually happening when energy levels with macroscopic occupancy "cross" as a function of breaking the symmetry between the trap wells. It is seen that only a few particles are actually active in the level crossing. Recent experiments in super-conductive Josephson loops showing tunneling of macroscopic quantum states are easily included in this picture.

# 1. INTRODUCTION

In 1924, even before the advent of modern quantum mechanics, Bose and Einstein posited the possibility of particles, now called Bosons, eschewing the exclusion principle, and Einstein (1925) pointed out the possibility of an ideal gas of such particles achieving a density wherein some *would not participate in the usual thermal agitation*. A composite particle, such as a neutral atom, will be a Boson if it contains an even number of total Fermions, and interacts with its environment in such a manner that its composite nature is not apparent. Thus, for example $^{23}$Na is a Boson, with 23 neutrons and protons, and 11 electrons; while a mass 22 or 24 isotope of sodium would be a composite Fermion. In modern terms Einstein's criterion for such a *condensation*, referred to as a Bose-Einstein condensate or BEC, is that the relative deBroglie wavelength would exceed the mean separation between atoms. Approximate ideality requires that both of these length scales be large compared to an atomic size. To condense an actual gas of Bosonic atoms, such as $^1$H, $^7$Li, $^{23}$Na, or $^{85,87}$Rb in a magnetic trap, held via the magnetic moment of the outer unpaired electron, temperatures of the order of tens of nano-Kelvin are required. Such conditions were first achieved, and clear evidence of a quantum phase transition to the condensate observed, in the laboratory in 1995. Good overviews of the pre-and post- 1995 state of experiments and theory appear in Griffin, Snoke, and Stringari (1995) and Pethick and Smith (2002), respectively. Recent theoretical overviews are Dalfovo et. al., (1999), and Leggett (2001).

The Griffin, Snoke, and Stringari (1995) collection of articles contains extended discussions of condensates in more complex condensed matter systems: superfluids and superconductors are Bose condensates. Tinkham (1996) discusses superconductors; and, Tilley and Tilley (1990) overview both superfluids and superconductors. Both, of course, had been observed well before those in the gas phase. What makes the gaseous condensates distinctly different from those of traditional condensed matter physics is that in the gaseous condensate: i) number densities are under experimental control over many orders of magnitude; ii) via optical pumping between hyperfine states, the number of components in a condensate is under experimental control; iii) the effective interaction between condensed atoms is fully tunable in many systems, even with respect to sign; iv) optical manipulations of the original single well magnetic traps allows creation of two or many well condensates, with the tunneling of particles between them also tunable; v) the condensates are *naked*, that is not hidden inside a solid or liquid, a photograph or other image of a gaseous condensate is an image of the condensate particle density itself. The gaseous BEC thus offers an exceptional playground for theory and experiment, and thus a test bed for understanding and testing some of the basic underpinnings of condensed matter quantum physics.

What does all of this have to do with quantum chemistry? We take a condensate in a single well potential to be the analog of an atom, with the trap playing the role of the attractive nucleus, and the electrons being replaced by the condensate atoms, which being uncharged can easily be of variable number, and ask for the ground state to be described by a fully symmetric Hartree wave function. We then enquire as to the importance of correlations. These are typically found to be small in three-dimensional single well traps, but important as one or two of the dimensions are reduced beyond a critical length scale, see, for example, Girardeau et. al. (2001), and references therein.

However, if a single well condensate is the analog of an atom, then multi-well condensates are molecules! Once molecules are introduced, any self-respecting quantum chemist will know that correlations must be properly included if dissociation is to be described. Experimentally in a gaseous BEC such a dissociative process is called *condensate fragmentation*, and may be achieved by raising the tunneling barrier in a double (or multi-) well condensate as discussed in Orzel et. al. (2001) and Greiner et. al. (2002). Extended multi-well fragmentation is a "Mott" transition, familiar in the theory of insulator-conductor transitions in the band theory of electrons in solids, as discussed in, for example, Ashcroft and Mermin (1976).

More recently neutral Fermionic atoms have been condensed in the gas phase, DeMarco and Jin, (1999, 2002), offering the possibility of model atoms and molecules with arbitrary numbers of wells and arbitrary numbers of "uncharged" analogs of electrons. When correlations become important in either the Bose or Fermi cases, a clear opportunity for quantum chemists arises. Per-Olov would be delighted!

We make no attempt to give a complete overview of this material here. Rather, in the style of traditional quantum chemistry we assume that a proper zero order description of the ground state is to simply put all of the Bosons in a single configuration of atomic or molecular orbitals (MOs), and then use the appropriate mean-field or configuration interaction (CI) corrections, for example see McWeeny and Sutcliffe (1969), to achieve the necessary level of description. Once CI is introduced, a natural question is as to what diagnostics are to be used to allow us to describe "what the correlated particles" are doing in terms of an orbital picture. A most natural and convenient manner in which to do this is to use the Bosonic equivalent of Löwdin's natural orbitals (NOs) and their corresponding occupation numbers, Löwdin (1955).

To illustrate the utility of the NO-MO analysis we consider the simple problem of a gaseous BEC in a symmetric double well trap. This "two mode" BEC problem has been discussed many times before in quite different languages, see, for example: Milburn et. al. (1997), Cirac et. al. (1998), Steele and Collett (1998), Gordon et. al. (1999), Raghavan et. al. (1999), Spekkens and Sipe (1999), Franzosi and Penna (2001).

Three problems will be of interest here:

1) understanding the multi-state CI (Born-Oppenheimer like) state correlation diagram as a function of tunneling strength in an MO picture. Alternative, and quite different, discussions of this problem appear in Steele and Collett (1998), and Franzosi and Penna (2001).

2) the description of condensate fragmentation in the NO-MO framework both for the ground and highly excited states; and,

3) the understanding of what particles are actually involved in the tunneling of macroscopic quantum states as these become near degenerate and are only weakly coupled via a tunneling barrier. We will see that the avoided crossings of macroscopic tunneling states can involve a very small number of active particles.

In the course of generating the full correlation diagram, looking at all of the energy levels resulting from a full CI (complete CI in a restricted basis) as a function of barrier height between the double wells, we find immediate evidence for a collective and sharp macroscopic quantum transition. This observation, although made in a novel manner, is not at all new: we are immediately taken back to P. W. Andersons's interpretation of the Josephson effect as a physical pendulum; Anderson (1964, 1966, and 1984). The physical pendulum has two distinctly differing types of motion: oscillations about a single minimum, and hindered rotation. Projection of our CI wave functions into the proper semi-classical phase space, Husimi (1940), shows that these two types of motion are precisely what is implied in the various types of quantum mechanical excited states of the double well condensate. Where hindered or free rotation is at hand both energies and occupation numbers are (exactly or nearly) doubly degenerate, as the rotation may be in either of two directions, and a simple linear combination of the correlated wavefunctions gives fully localized quantum states, and NOs. We thus interpret the correlation diagram in terms of an MO to AO localization process.

The outline of the paper is a follows: in Section 2 the Hartree theory of a BEC localized in a single well is introduced, and its low energy excitations briefly discussed. In particular, it is noted that typical low energy excitations are "collective" motions of all particles, rather than the atomic/molecular picture of "single particle excitations" from a single ground state configuration. The reason for the lack of low-lying single particle excitations is discussed following Huang and Yang (1957). The single particle excitations will, perhaps surprisingly, and pleasingly to quantum chemists who might wonder where they appear, reappear in the excitation spectrum of the double well condensate, and in the description of macroscopic quantum tunneling. A simple highly restricted CI model for a

BEC in a double well potential is then introduced following the work of Spekkens and Sipe (1999), who focus on fragmentation of the ground state. The Spekkens-Sipe Hamiltonian is seen to be easily parameterized, and to then simplify to a Bose-Hubbard site-to-site tunneling model, see, for example Fisher et.al. (1989), used directly in many of the previous studies of two-mode BECs, and indicating the generality of the results obtained here. Finally, in this introductory section, the first order reduced density matrix (1-matrix) is introduced, and factored. This is first done within the framework of the Onsager-Penrose, Penrose (1951), Penrose and Onsager (1956), *ansatz* of ODLRO, or "Off Diagonal Long Range Order." Mathematical implementation of the ODLRO *ansatz* requires introduction of phase-number coherent states, Anderson (1966), Loudon (1973), if accomplished directly as a factorization in $2^{nd}$ quantization. These coherent states do not correspond to states with a fixed particle number, but are introduced as they will prove useful in the later analysis. As an alternative to the use of coherent states to obtain the ODLRO factorization, the Löwdin NO-occupation number analysis is introduced; see, Löwdin (1955), McWeeny (1959), Davidson, (1976). In Section 3 an overview of numerical results for the full CI spectrum of energy states of the double well BEC is given as a function of barrier height separating the two wells. Condensate fragmentation is observed. The correlation diagram, and occupation number analysis, shows immediate and clear evidence for a macroscopic quantum phase transition, which is the crossover from a delocalized MO picture to degenerate pairs of states which may be easily combined to give a localized orbital description.

The interpretation of this crossover is discussed in Section 4, where contact with the early work of Anderson (1964, 1966) and his interpretation of the Josephson effect as a "physical pendulum" is made. Using an appropriate coherent state basis, a Husimi projection, Husimi (1940), Martens (1989), from individual stationary states for the N particle system is shown to give precisely the semi-classical behavior predicted by Anderson, even for N as small as 20 to 40 particles. The description of the fragmented "rotor" states in terms of localized CI states and corresponding localized NOs is then given. In Section 5 the problem of macroscopic quantum tunneling, recently observed in superconducting current loops, i.e. see van der Wal et. al., (2000), Freidman et. al. (2000), and Yu et. al. (2002), is considered. The symmetry of the Spekkens-Sipe double well is broken by addition of an additional parameter: as a function of this parameter, rotor state degeneracy is broken, and the quantum tunneling splittings studied. Near the crossover region these tunneling splittings are observable. The appropriate NO analysis shows, however, that actually only a few (usually one to three, but variable according to the tuning and magnitude of the splitting) particles, out of many hundreds modeled theoretically, or millions in actual experiments, are actually "moving" from one MO to another in the crossing process. In this situation "few body" excitations actually become the dominant excitation mechanism, in strong contrast to the usual picture of superfluid excitations being collective modes. This also immediately clarifies why a single microwave photon can couple such macroscopic states, an issue not addressed clearly in the experimental papers above. Section 6 contains conclusions and prospects for further work.

## 2. A Gaseous BEC in a Double Well Potential: Correlations and Their Analysis

In this Section the mean-field equations for a BEC in a single potential well are briefly derived from a wavefunction point of view, and the Fermi-contact interaction is introduced (Section 2.1). This results in the Gross-Pitaevskii (GP), or non-linear Schrödinger equation (NLSE); see Gross (1961), Pitaevskii (1961), Sulem and Sulem (1999), respectively. The single configuration GP level description encompasses much of the observed behavior of trapped BECs. Densities, low-lying collective excitations, e.g. Ruprecht et. al. (1995), Edwards et. al, (1996) and even strongly localized non-linear excitations such as solitons, Reinhardt and Clark (1997), Reinhardt (1998, and references therein) and vortices, Gross (1961), Pitaevskii (1961) are collective modes, in that all N particles remain in an identical single particle orbital. Notably absent are low energy "single particle" excitations of the type common in atoms and molecules where the usual optical transitions take a single electron from one orbital to another. The reason for absence of such single particle excitations appears in first order Rayleigh-Schrödinger perturbation theory, as exposed by Huang and Yang (1957). It may be interpreted as a Bosonic amplification effect for particles to remain in the same single particle state, once that single state has "macroscopic occupancy". This restriction is immediately removed if instead of a single condensate, we have two, or more, weakly coupled condensates. A single particle may then easily jump from one already macroscopically occupied state to another. Tunneling between macroscopically occupied states is a facile single particle process controlled by a one particle tunneling matrix element. We will not, then, be surprised, to see that the NO analysis of excited states of a double well condensate shows that states typically differ by moving small numbers of particles from one NO to another. In Section 2.3 the model Hamiltonian used herein is introduced, and it's relationship to the Bose-Hubbard model of condensed matter physics made clear. Finally, the 1-matrix is introduced in Section 2.4, and factored: i) by fiat; and, ii) more sensibly to a quantum chemist, by following the lead of P.-O. Löwdin, by introduction of a natural orbital analysis, Löwdin (1955).

## 2.1 Mean Field Theory for a Single Well BEC

The usual derivation of the condensed matter mean field theory follows from factorization of the ground state expectation value of the second quantized Heisenberg equation of motion for a Bosonic field operator. See, for example, Griffin (1996), Fetter and Walecka (1971). Quantum chemists will not be surprised that identical mean field equations may be derived from the *ansatz* that the ground state wavefunction for an N Boson ground state consists of the fully symmetric product of N identical one particle orbitals (MOs in what follows), and then minimizing the energy of this single "perminantal" wavefunction with respect to its single orbital. This is simply the Bosonic analog of, for example, the derivation of the Roothaan restricted Hartree-Fock equations for a closed shell atom or molecule from a single "determinantal" wavefunction, Roothaan (1952). The resulting Hartree (i.e. no exchange) equation is

$$\left(-\frac{\hbar^2}{2m}\nabla_r^2 + V^{trap}(r)\right)\phi(r) + \sum_{i=1}^{N-1}\left[\int \phi_i^*(r')v(r',r)\phi_i(r')d^3r'\right]\phi(r)$$
$$= \mu\phi(r) \qquad (2.1)$$

where $v(r_i, r_j)$ is the pair potential. Here, a physical assumption has been made. For non-relativistic electrons, for example, there *is* only a spatial pair interaction $-1/r_{ij}$. This is not true for atoms in a condensate. For a gas of cold atoms it must be assumed that the density is low enough so that all three-body effects, including three body contributions to the potential energy, may be neglected in writing Eqn. (2.1). This is consistent with the existence of the condensate itself: if three body interactions become important, actual diatomic molecules or clusters may form, and be lost from the condensate.

Given that the condition for formation of a BEC is that local relative deBroglie wavelengths are long compared to inter-particle distances, the local interaction between any pair of atoms is dominated by the s-wave interaction. Following Fermi, the pair interaction may then be replaced by the contact, or point, interaction, long familiar as the Fermi contact interaction in the description of an s-state electron interacting with an atomic nucleus (see, for example, Flygare, 1978). The contact interaction "pseudo-potential" replacement is

$$v(r_i, r_j) \rightarrow \frac{4\pi\alpha_s\hbar^2}{m}\delta^3(r_i - r_j) \equiv g\delta^3(r_i - r_j) \qquad (2.2)$$

and is thoroughly discussed by Huang, Lee, and Yang (Huang and Lee, 1957), and Lee et. al. (1957), in the context of the perturbation theory of an extended N Boson system, where the ground state orbital is taken as a k = 0 plane wave, namely a constant. The constant $\alpha_s$ in Eqn (2.2) is the s-wave scattering length, and is taken so that a Born approximation to the scattering amplitude gives the "exact" s-wave scattering amplitude. This is thus the lowest order part of a renormalization process, by which a hard core repulsive and short range attractive Born-Oppenheimer molecular pair potential of considerable complexity is replaced by a regularized potential, allowing plane wave matrix elements to be taken in a simple manner. This is a matter of some subtlety, see Huang (1957), above, and also Leggett (2001). Accepting the replacement of the pair potential by the contact interaction of Eqn (2.2), which is certainly fully empirically justified, the Hartree mean field equation for the ground state single particle orbital, $\phi$, becomes:

$$\left(-\frac{\hbar^2}{2m}\nabla_r^2 + V^{trap}(r) + g(N-1)|\phi(r)|^2\right)\phi(r) = \mu\phi(r), \qquad (2.3)$$

as all of the $\phi_i$ are the same for the ground state configuration. Hartree self-consistency follows from the requirement that the same single particle orbital $\phi(r)$ sits inside the effective potential, $g|\phi(r)|^2$, and also solves the mean field eigenvalue equation (2.3). The eigenvalue $\mu$ is, as usual, a chemical potential, or "ionization" energy, i.e. Koopman's Theorem holds. In an extended large N system of neutral atoms, the energy to add or subtract a single particle will be the same, and both equal to $\mu$. Eqn. (2.3) is usually referred to as the non-linear Schrödinger equation (NLSE) or the Gross-Pitaevskii (GP) equation, from its original use in condensed matter theory, see again, Gross (1961), Piteavskii (1961), Sulem and Sulem (1999). Hartree (1928) could have written down Eqn. (2.3) at once, if he had been considering Bosons, and had known to employ the pseudo-potential. Bravery, hubris, and spectacular empirical agreement with many classes of experiment suggest the time dependent generalization

$$i\hbar\frac{\partial}{\partial t}\phi(r,t) = \left(-\frac{\hbar^2}{2m}\nabla_r^2 + V^{trap}(r) + g(N-1)|\phi(r,t)|^2\right)\phi(r,t). \qquad (2.4)$$

Eqn. (2.4) has been successfully used to describe the dynamics of phonons, where the time dependent GP equation gives energies equivalent to those of the RPA, Ruprecht et. al. (1995), Edwards et.al. (1996), Thouless (1961). It also well describes dark solitons, Reinhardt and Clark (1997), Reinhardt (1998), Denschlag et. al. (2000); bright solitons, Strecker et.al. (2002), Khaykovich et. al. (2002); and, soliton-vortex interactions, Anderson, B. P., et. al. (2001), and vortex-solitonic structures, Brand and Reinhardt (2002), in properly confined condensates. Leggett (2001) has correctly noted that these seemingly successful applications can take the GP theory way beyond its proven domain of validity. For example, Eqn. (2.4) does not even describe the decay of the excitations it seemingly describes well. Thus its usage is not fully theoretically justified, as the timescales over which it is valid are currently unknown. This, of course, has not prevented many successful applications. The use of the GP, or time dependent GP, equations *does* neglect correlated motions, and thus cannot correctly describe excitations wherein one or more particles "leave" the ground state configuration. Correlations do dominate the dynamics of 1-dimensional condensates, e.g. Girardeau (2001), and references therein. However, in 3 dimensions even creation of strongly localized "high energy" excitations, whether these are kinks in stationary states (Carr et. al., 2000a,b), or time evolving excitations such as solitons or vortices, see above, may not involve correlations, and may thus be described by a single configuration N-Boson wavefunction, wherein all N Bosons have an identical single particle wavefunction, $\phi(r,t)$. As long as this is the case, the time dependent Hartree theory of Eqn. (2.4) can be applied. Where, then, are the "single and double" excitations familiar to quantum chemists in the CI picture of correlations in the electronic structure of atoms and molecules, where they are needed for accurate determination of energetics, and also for proper description of dissociation? Such excitations also form the basis for discussion of the low energy shell model electronic spectroscopy of these same systems. A hint as to their absence is given by first order perturbation theory.

## 2.2 First Order Perturbation Theory for Single Particle Excitations

Huang and Yang (1957) introduced a first order perturbative description of a dilute, hard sphere Bose gas in a macroscopic uniform volume, *V*. That is, N Bosons in a box with periodic boundary conditions. This was soon followed by estimates of the $2^{nd}$ and $3^{rd}$ order corrections to the energy and density in the same uniform thermodynamic limit, Lee et. al., (1957), and Wu (1959). However, the simple first order result is of exceptional interest, in that, for a condensate of repulsive effective interaction, i.e. positive scattering length, $\alpha_s$, it is immediately seen that N-

Boson systems achieve low energy via multiple occupancy of a single orbital. Consider a uniform Bose system with zero order energies, $\varepsilon_i$, and corresponding occupation numbers, $n_i$. The zero order energy is

$$E_0 = \sum_{i=0}^{\infty} n_i \varepsilon_i \qquad (2.5)$$

where $N = \Sigma_i\, n_i$. The ground state zero order energy is thus given by the configuration $n_0 = N$, and $n_i = 0$ for all $i > 0$, and is thus $E_0 = N\varepsilon_0$. Huang and Yang (1957) found that the first order correction to this unperturbed energy for a configuration with occupation numbers $n_i$ is

$$E_1 = \left(\frac{g}{V}\right)\left(N^2 - \frac{1}{2}N - \frac{1}{2}\sum_{i=0}^{\infty} n_i^2\right) . \qquad (2.6)$$

For the ground state this immediately reduces to (for $n_0 = N$)

$$E_1^{(0)} = \left(\frac{g}{V}\right)\left(\frac{N(N-1)}{2}\right) \qquad (2.7)$$

which is simply the number of pairs of Bosons, N(N-1)/2, times the effective pair interaction strength. At first glance the expression of Eqn. (2.6) seems hard to interpret; however, consider the "first" excited configuration, $n_0 = (N-1)$, and $n_1 = 1$. Eqn. (2.6) may then be rearranged as

$$\left(\frac{g}{V}\right)\left(\frac{N(N-1)}{2}\right) + \left(\frac{g}{V}\right)(N-1). \qquad (2.8)$$

Comparing Eqns. (2.7) and (2.8) it is evident that the energetic cost of the single particle excitation is

$$\Delta = \left(\frac{g}{V}\right)(N-1). \qquad (2.9)$$

As $(N-1) \sim N$ for a macroscopic, or even mesoscopic, system (typical laboratory gaseous BECs contain $10^4$ to $10^{11}$ particles) Eqn. (2.9) shows that there is a *macroscopic* cost for creating a single particle excitation. This situation does not arise for electrons, as, due to the exclusion principle, no Fermionic orbitals have such macroscopic occupancy. This is the origin of the "force" leading towards the Bose condensation. Restating this Huang/Yang idea: dilute, repulsive ($\alpha_s > 0$), Bose condensates condense into a single particle state space, these being plane waves for an extended uniform system, or the ground state orbital, $\phi(r)$, in a trap. Conversely, attractive condensates ($\alpha_s < 0$) condense in configuration space, where the NLSE is actually unstable with respect to collapse of the mean field wavefunction for large enough N. See, Kosmatov, et. al. (1991), Sulem and Sulem (1999), Sackett et. al. (1999), Roberts et. al. (2001), for discussions of the theory and the current experimental situation for the instability of attractive condensates. We restrict our attention to repulsive condensates in what follows. Interestingly, this local stability of repulsive condensates with respect to excitation of a single particle from a single macroscopically occupied configuration does not only apply to the ground state configuration. Macroscopic occupancy of an excited state is also locally stable with respect to single particle excitations, or de-excitations! Low lying excitations for a repulsive condensate are thus "collective" sloshings of the density, wherein all N particles stay in the same, now time dependent, orbital $\phi(r,t)$. This is the Bogoliubov picture of phonon excitations, see for example, Fetter and Walecka (1971), Leggett (2001). This, also, may be part of the explanation of the success of the time dependent GP description of highly energetic and non-linear phenomenon such as solitons, vortices and their interactions, a success surprising to traditional condensed matter physicists (again, e.g. Leggett (2001), page 352).

The argument above should not, however, be construed to imply that BECs are weakly interacting: the same factor of N in Eqn (2.9) implying a high cost of single particle excitations also appears in the effective potential of the GP equation (2.4): mean field effects are often enormous, and, for example can completely out weigh the kinetic energy. In this, the Thomas-Fermi limit, Eqn. (2.4) may be simply solved for the ground state particle density:

$$\rho(r) = N |\phi(r)|^2 = (\mu - V^{trap}(r))/g, \quad (2.10)$$

where $(\mu - V^{trap}(r)) > 0$. This is, in fact, often an excellent approximation to actual ground state densities of trapped (repulsive) BECs.

An interesting question for quantum chemists thus arises: When will correlations involving excitations of a small number of particles from the ground state configuration occur and be observable? The answer is clear: such excitations are needed to properly describe the division, or fragmentation, of a condensate into two (or more) parts. In this case the mean field MO picture does not allow localization of the particles in the separated pieces of the condensate. Said another way, just as in the theory of electronic structure, the MO picture does not allow molecules to dissociate to their atomic ground states: correlation is required.

Additionally, for a BEC trapped in a double well, transfer of a single particle from one well to another can be a low energy event, simply because the particle is transferred from one macroscopically occupied state, say in the left well, to another already macroscopically occupied state in the neighboring right well.  The perturbative argument of Huang and Yang, seemingly forbidding single particle excitations (or, more generally, few body excitations from the initial configuration), no longer applies as we have both a macroscopic energy cost, and an equivalent macroscopic re-capture of energy, as a single particle is transferred from one condensate of identical particles to another.  We now introduce a simple model for such a double well condensate.

## 2.3 A Simple Model for a BEC in a Double Well

Many recent papers have treated the problem of a double well condensate.  For example: Milburn et. al. (1997), Cirac et. al. (1998), Gordon et. al. (1999), Raghavan et. al. (1999), Spekkens and Sipe (1999), Franzosi and Penna (2001), Leggett (2001, for an overview), to name only a few.  We introduce the model of Spekkens and Sipe (1999) here. Spekkens and Sipe consider the problem of fragmentation of a condensate ground state as a barrier height is raised. Our analysis goes well beyond this, but begins with the same model Hamiltonian. Approximation, and parameteriztion, of the Spekkens-Sipe model gives rise to the Bosonic Hubbard Hamiltonian (Fisher et. al. 1989), which may be used for double and multi-well tunneling processes.

Figure 1 shows a typical symmetric double well potential, V(x), in a one dimensional co-ordinate space, and we will use this model to typify the correlations which arise when a condensate is trapped in such a situation. Symmetry requires that the only two NOs are the symmetric and anti-symmetric linear combinations of the localized AOs.  Neglecting mean field effects, which *is* a serious approximation but we focus on correlation effects here, let us assume that if a single particle is in the left well its localized Schrödinger wavefunction (AO) is $\phi_L(r)$, and if on the right, $\phi_R(r)$.

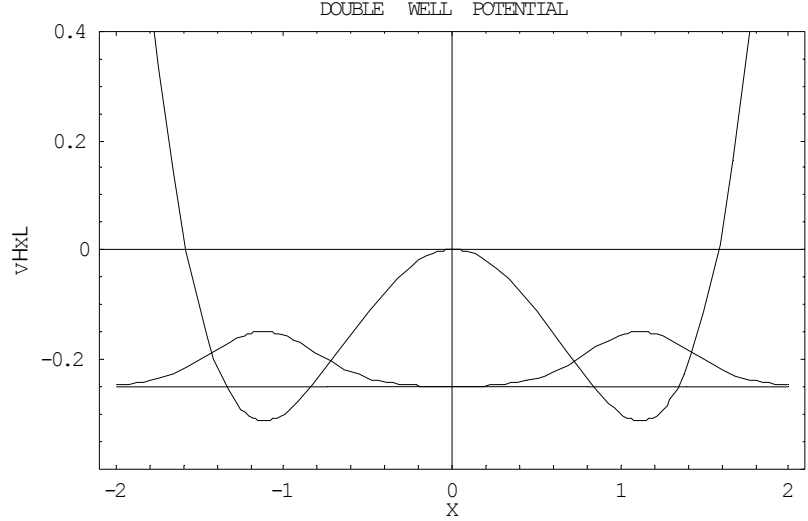

*Figure 1.* A double well potential, V(x), which might contain a single particle, or many particles. Shown also, are localized "atomic like" (AOs) orbitals on the right, $\phi_R(x)$, and the left, $\phi_L(x)$. Molecular like orbitals may be found as linear combinations of these.

As it is convenient to work in the language of second quantization, let the Fock state $|N,M\rangle$ denote the state with N particles on the left, and M on the right of the barrier, in the single particle states $\phi_L(r)$ and $\phi_R(r)$, respectively. Denoting the corresponding single particle creation-annihilation operators as $C^\dagger_L, C_L$ and $C^\dagger_R, C_R$, the Spekkens-Sipe Hamiltonian is:

$$H = \varepsilon_{RR}\hat{N}_R + \varepsilon_{LL}\hat{N}_L + \left(\varepsilon_{LR} + gT_1(\hat{N}-1)\right)(C^\dagger_L C_R + C^\dagger_R C_L)$$

$$+ g\frac{T_0}{2}(\hat{N}_L^2 + \hat{N}_R^2 - \hat{N})$$

$$+ g\frac{T_2}{2}(C^\dagger_L C^\dagger_L C_R C_R + C^\dagger_R C^\dagger_R C_L C_L + 4\hat{N}_L \hat{N}_R) \quad (2.11)$$

where $\hat{N}_L = C^\dagger_L C_L$; $\hat{N}_R = C^\dagger_R C_R$; $\hat{N} = \hat{N}_L + \hat{N}_R$ and

$$\varepsilon_{ij} = \int \phi_i(r)\left(-\frac{\hbar^2}{2m}\nabla^2 + V^{trap}(r)\right)\phi_j(r)d^3r \quad \text{where} \quad i,j = R, L$$

$$T_0 = \int \phi_{L,R}^4(r)d^3r; \quad T_1 = \int \phi_L^3(r)\phi_R(r)d^3r$$

$$T_2 = \int \phi_L^2(r)\phi_R^2(r)d^3r \quad (2.12)$$

These matrix elements are easily interpreted. The $\varepsilon_{LL}$ and $\varepsilon_{RR}$ are the energies of a single particle in the left and right wells, each is multiplied by the number operator $N_{L,R}$ counting the particles in each well. $\varepsilon_{RL}$ is a single particle tunneling amplitude, it appears as a coefficient of operators which allow a single particle to hop from one well to the other. $T_0$ is the non-linear mean-field energy associated with particles in a single well. $T_{1,2}$ are non-linear tunneling matrix elements; both are proportional to the coupling constant g, and both enter with quadratic functions of particle numbers, as in Eqn. (2.7). To allow investigation of the effect of the change in barrier height, we have simply parameterized the matrix elements as follows for the symmetric potential case:

$$\varepsilon_{RR} = \varepsilon_{LL} = T_0 = 1; \quad \varepsilon_{LR} = T_1 = -e^{-\alpha}; \quad T_2 = e^{-2\alpha} \quad (2.13)$$

This parameterization allows a simple study of continuous change in the linear and non-linear tunneling via variation of the single parameter $\alpha$, not to be confused with the scattering length, $\alpha_s$, which is part of the effective coupling constant "g". Large $\alpha$ implies small tunneling, leading to fragmentation.

In Section 3 we present results of carrying out full CI in the restricted N particle basis of Fock states |N-n, n>, n = 0,1,…N allowing the possibility of all particles being in either well, and all intermediate arrangements. This is carried out as a function of $\alpha$, yielding a correlation diagram. Such a full CI is, as usual, size consistent, and may thus be applied to large numbers of particles. The simplicity of the two state double well model is that an N particle problem results in a (N+1) x (N+1) dimensional CI matrix, with eigenvector $\{c^i\}$ for the i$^{th}$ eigenvalue of $H(\alpha)$. The occupation representation eigenfunctions are then of the form

$$|\Psi_i\rangle = \sum_{n=0}^{N} c_n^{(i)} |N-n,n\rangle . \quad (2.14)$$

Before giving results of computations with in this model, we note that for large N, the $\varepsilon_{LR}$ term may be neglected, and if we also ignore the terms $T_2$ which scales as $e^{-2\alpha}$, rather than $e^{-\alpha}$, the model reduces to

$$H = \hat{N} + g\frac{T_0}{2}(\hat{N}_L^2 + \hat{N}_R^2 - \hat{N}) + gNT_1(C_L^\dagger C_R + C_R^\dagger C_L), \quad (2.15)$$

which is similar in form to a two well Bosonic-Hubbard model (Fisher, 1989), but differs in that the there is a factor of "N" in the $T_1$ tunneling term. The usual Bose-Hubbard model contains only the linear Schrödinger-like, tunneling term $\varepsilon_{LR}$. However for fixed N, the two models are simply related by a scaling of parameters, with perhaps an additive constant. The R,L labels may also be extended to an arbitrary number of wells, leading to the widespread use of this model in discussions of BECs in multi-well optical traps.

## 2.4 Analysis of Correlated Wavefunctions: The 1-Matrix, ODLRO, Coherent States and Natural Orbitals

The key tool in our analysis of the correlated wavefunctions, obtained from the full CI calculations for the double well BEC as modeled in Section 2.3, is the first order reduced density matrix, or 1-matrix. The 1-matrix has played a key role in the theory of super-conductivity, as well as in quantum chemistry. P.-O. Löwdin played a seminal role in this latter development. In second quantized notation, for example Fetter and Walecka (1971), the 1-matrix for an N body ground state (approximate or exact) |N,0⟩ is defined as

$$\gamma(1,1') = \langle N,0 | \psi^\dagger(1)\psi(1') | N,0 \rangle , \qquad (2.16)$$

where $\psi^\dagger(1)$, $\psi(1')$ are the Bosonic field operators, $\psi^\dagger(1)$ creating a particle at space-spin point 1, and $\psi(1')$ destroying a particle at 1'. The diagonal element, $\gamma(1,1)$, is, of course, the particle density $\rho(1)$. $\gamma(1,1')$ is normally a non-separable function of the two space-spin co-ordinates 1,1'. Penrose (1951), and Penrose and Onsager (1956), proposed that a key descriptor and indication of a transition to the superfluid or superconducting state would be the factorization:

$$\gamma(1,1') = f(1)f(1') \qquad (2.17)$$

for a suitable function $f(1)$. In the superfluid literature a 1-matrix for which this property holds is said to have the property off-diagonal long range order, or ODLRO, namely that the ground state is uniform in the sense that there is no correlation between points at which the system is probed by the operator $\psi^\dagger(1)\psi(1')$. We will see below, following Löwdin, that this is simply a statement of macroscopic occupancy of a single quantum state by the vast majority of particles.

But, we first ask, what is the function $f(1)$? P. W. Anderson (1966) suggested that ODLRO resulted from the factorization

$$\langle \psi^\dagger(1)\psi(1') \rangle = \langle \psi^\dagger(1) \rangle \langle \psi(1') \rangle . \qquad (2.18)$$

In Eqn. (2.18) it is immediately evident that the expectation value $\langle,\rangle$ cannot be taken with respect to the N particle ground state $|N,0\rangle$, as this is a state with precisely N particles, and the operators $\psi^\dagger$, $\psi$ only have matrix elements between states which have particle numbers differing by 1. Anderson thus suggested that the appropriate expectation value to take to effect the possible ODLRO factorization would be a non-number conserving *coherent state* of a form such as :

$$|\Theta\rangle = \sum_N a_N |N,0\rangle, \qquad a_N \cong e^{i\theta N} Exp\left(-\frac{(N-\overline{N})^2}{(\Delta N)^2}\right). \qquad (2.19)$$

Use of such states has many positive attributes: in diagrammatic perturbation theory a version of Wick's theorem exists for coherent states, but not for Bosonic Fock states, see Fetter and Walecka (1971). Use of coherent states allows a distinct phase to be associated with the condensate wave function, now defined as

$$\phi(r) = \langle \psi(r) \rangle = \sqrt{\rho(r)} e^{i\theta} \qquad (2.20)$$

and identified with the Landau *complex order parameter* empirically used in the phenomenological theory of superfluids, see, for example, Tilley and Tilley (1990). The phase of the condensate wave function is of exceptional importance: phase gradients drive supercurrents, Anderson (1964,1966); and the phase offset, $\Delta\theta$, between two weakly coupled superfluids drives a current proportional to $\sin(\Delta\theta)$. This latter is the Josephson effect of superconductivity. See Josephson (1962), Anderson (1964, 1966), Tinkham 1996). $\Delta\theta$ also controls the velocity of solitons in the BEC: Reinhardt and Clark (1997). We will make use of coherent states, and a projection into a phase space defined by the conjugate variables (N,θ) in Section 4. The fact that the particle number and phase are conjugate quantum mechanical variables implies an uncertainty relationship, the meaning of which will also be illustrated in Section 4.

In the thermodynamic limit $N \to \infty$, at constant density, $\rho = N/V$, the particle number uncertainty to maintain a fixed phase is vanishingly small compared to the overall number of particles, and the use of the Anderson coherent states, and the use of Eqns. (2.19 and 2.20) is perfectly sensible.

The fact that a condensate has some definite phase (out of all possible phases) is then considered a gauge symmetry breaking, analogous to the broken symmetry as a magnet orients its magnetic moment in one particular direction, rather than another. However, it does seem un-natural to assume that in a laboratory experiment with, say, 11,929 particles in a BEC in a magnetic trap, that we should consider the system *not* to have a fixed number of particles, or that even the wavefunction for a single particle might not have a perfectly definite phase, even if the physics is dependent on its relative, rather than absolute value. These statements are not universally accepted, at least partly due to semantic confusion, and the shared view of many condensed matter theorists that no aspects of the gaseous BEC can be simply described by direct use of a wavefunction. To paraphrase Leggett (2001, page 317), see also Reinhardt (1998, pages 281-4): *It should be emphasized that there are no circumstances in which Eqns. (2.19-20) are the physically correct description of the system, or even a part of it…However, the reader should be warned that this opinion is controversial, and there are even those who feel that Eqns. (2.19-20) are not only a possible, but the only legitimate, definition of the order parameter.*

With so much seemingly at stake, it is refreshing to note that if we follow Löwdin (1955), rather than Anderson (1964, 1966), a very simple interpretation of ODLRO for a fixed number of particles emerges immediately. The 1-matrix, considered as an integral kernel

$$\int \gamma(1',1)\varphi_i(1')d1' = \eta_i \varphi_i(1) \qquad (2.21)$$

generates a complete set of orthogonal natural orbitals, $\varphi_i$, (NOs hereafter) and corresponding occupation numbers, $\eta_i$. In terms of these

$$\gamma(1,1') = \sum \eta_i \varphi_i^*(1) \varphi_i(1'). \qquad (2.22)$$

The natural orbitals, if used in a CI calculation, provide optimal convergence for a one-particle basis. The occupation numbers $\eta_i$ tell us how important a particular NO is in a non-separable correlated wavefunction. The usual normalization of $\gamma(1,1')$ is such that

$$\sum \eta_i = N. \qquad (2.23)$$

For a Bosonic system a single quantum state may have macroscopic occupancy, and so, unlike the Fermi case discussed by Löwdin, where the spin NO occupation numbers are restricted, $0 \leq \eta_i \leq 1$, for Bosons this is replaced by $0 \leq \eta_i \leq N$, and only restricted by the fixed total number of particles, N, via Eqn. (2.23). Should a single occupation number dominate the sum, we have

$$\gamma(1,1') = N\varphi_0^*(1)\varphi_0(1'), \qquad (2.24)$$

which immediately gives an interpretation of the ODLRO factorization. Namely, the system has chosen to put essentially all of the particles in a single orbital, which is now seen not to necessarily be homogeneous in space. This is both macroscopic occupancy of a single state and factorization of $\gamma(1,1`)$, but without the artifice of introducing a non-number preserving formalism. The condensate phase, $\theta$, is seemingly lost, but is actually recoverable as seen in Section 4, where the Anderson coherent states will be seen to provide a most valuable interpretive insight into the nature of the dynamics of the correlated double well states, even though number conservation is kept exactly for the overall system.

## 3. Energy Level Correlation Diagram, and Natural Orbital/Occupation Numbers for the BEC Double Well Problem

In this Section the energy correlation diagram resulting from the full CI treatment of the double well problem of Section 2.3 is presented. This is followed by a natural orbital and occupation number analysis. Individual wavefunctions typifying various regions from the correlation diagram are presented. It is seen that both localized and delocalized (fragmented) states can occur for a wide range of values of $\alpha$, which measures the tunneling strength. For excited fragmented states it is seen that proper linear combinations of CI eigenstates are fully localized, even for tunneling couplings large enough so that the ground state is still delocalized. A physical interpretation of this dual behavior is presented in Section 4.

### 3.1 An Energy Correlation Diagram Shows a Phase Transition

Choosing the parameters in the Spekkens-Sipe Hamiltonian, Section 2, as $\varepsilon_{LL} = \varepsilon_{RR} = T_0 = 1$; $g = +0.8$; $\varepsilon_{LR} = T_1 = -\exp(-\alpha)$; and, $T_2 = \exp(-2\alpha)$, Figure 2 shows all 21 eigenvalues for a system of 20 particles in the double well potential for $\alpha$ ranging from $\alpha = 0$ to $\alpha = 5$. The tunneling parameters, $\varepsilon_{LR}$ and $T_1$, allowing particles to move through the barrier, thus vary from 1 to 0.0067, taking us from a region where the particles are completely delocalized between the two wells, to a region where most excited states are fragmented, and, if localized, will have definite numbers of particles in each of the two wells.

The correlation diagram of Figure 2 shows a striking set of energy level merging and avoided crossings. Starting at about $\alpha = 1.8$ the highest two energy levels become visually degenerate. As $\alpha$ continues to increase, corresponding to raising the barrier between the wells, subsequent pairs of levels do the same. In the limit of even larger $\alpha$, only the ground state of the repulsive double well condensate remains non-degenerate, and in that ground state 10 particles are in the state $\phi_R$ and 10 in $\phi_L$: the ground state is thus completely fragmented. Spekkens and Sipe (1999) discuss this limit for the ground state, with an identical conclusion. We focus on the overall picture of all states, and will attempt to understand the origins of the crossover ridge separating the non-degenerate delocalized states on the left of the diagram, from the doubly degenerate pairs on the right.

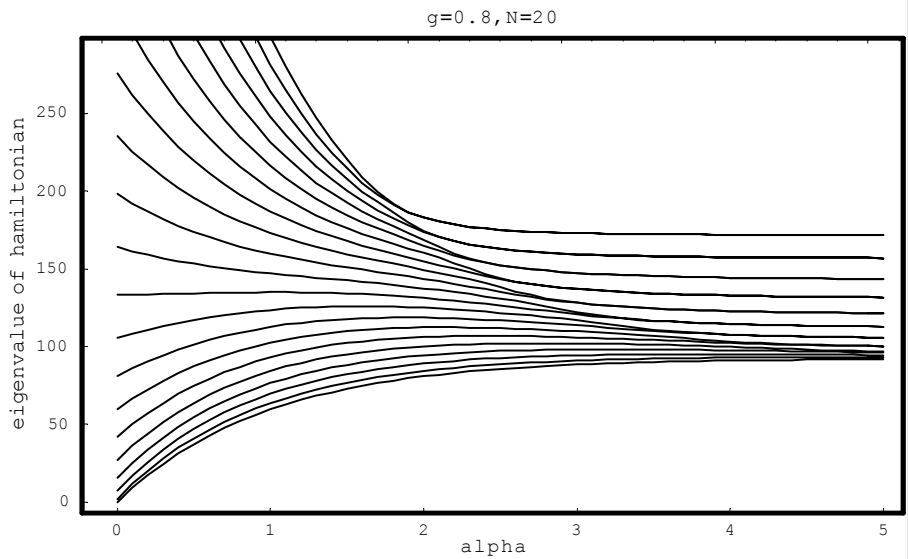

*Figure 2.* Correlation diagram for all energy levels for 20 Bose particles in a symmetric double well potential, as discussed in Section 2.3. Tunneling is controlled by the parameter $\alpha$, as $T_1 = -\exp(-\alpha)$, and thus tunneling is strongly allowed at the left, and mostly quenched on the right of the diagram. The most prominent feature of the diagram is the ridge of avoided crossings, and the onset of level degeneracies.

Our analysis of this correlation diagram and the structure of excited state wavefunctions is quite different from that discussed in the work of Steel and Collett (1998), and Franzosi and Penna (2001), who use a transformation to an angular momentum representation, as do Milburn et.al. (1997), in their analyses of wave functions of the form of Eqns. (2.14, 3.1) in discussions of the double well correlation problem. Neither of these groups introduce natural orbitals.

One can ask why we describe the small $\alpha$ orbitals as localized, and those for larger $\alpha$ as delocalized. Examination of a few of the coefficient vectors will make this distinction clear. Figure 3. shows the coefficients of the eigenvectors corresponding to the three lowest energy eigenvalues of the Spekkens-Sipe Hamiltonian, now for 40 particles. This number has been made larger simply to make the coefficient patterns "smoother" to the eye. What is plotted are the coefficients, $c^j_n$, of the number basis states $|N-n,n\rangle$ in the expansion for the $j^{th}$ eigenfunction of the Hamiltonian,

$$|\Psi_j\rangle = \sum_{n=0}^{N} c^j_n |N-n,n\rangle \qquad (3.1)$$

as a function of "n" the number of particles in the "right" well, and these are connected by lines to make their functional form more evident.

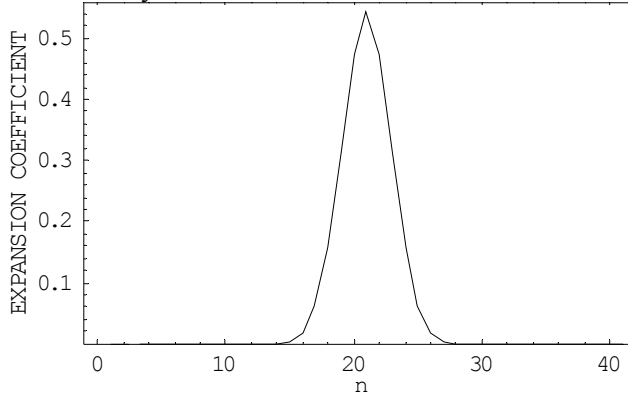

*Figure 3a*. The coefficients for the ground state of the double well BEC for N = 40, and $\alpha = 4$.

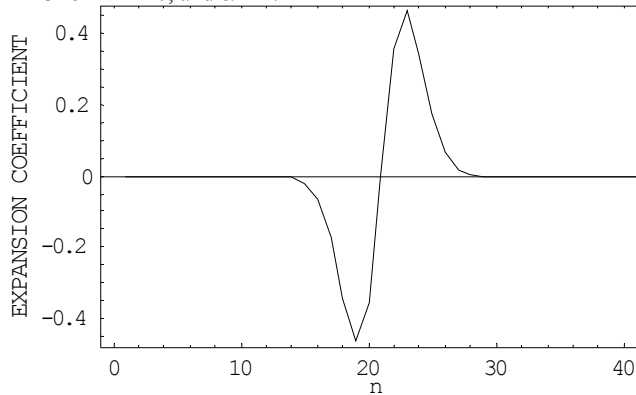

*Figure 3b*. The coefficients for the first excited state of the double well BEC for N = 40, and $\alpha = 4$.

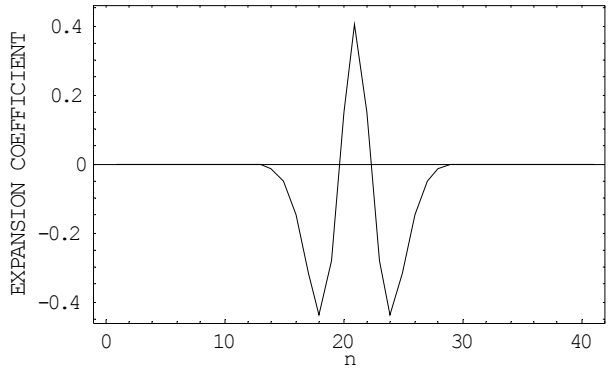

*Figure 3c.* The coefficients for the second excited state of the Double well BEC for N = 40, and $\alpha$ = 4.

These lists of coefficients were calculated for $\alpha$ = 4, and appear to be Harmonic Oscillator co-ordinate space wave functions. That this is not a coincidence is the subject of Section 4. Figure 4 shows similar lists of coefficients, again joined to help the eye follow their patterns, for two states on the other side of the ridge separating the doubly degenerate from the non-degenerate states. Note that these do not look like oscillator wave functions. Odd and even linear combinations of degenerate pairs give fully localized orbitals, one of which is shown in Figure 4c.

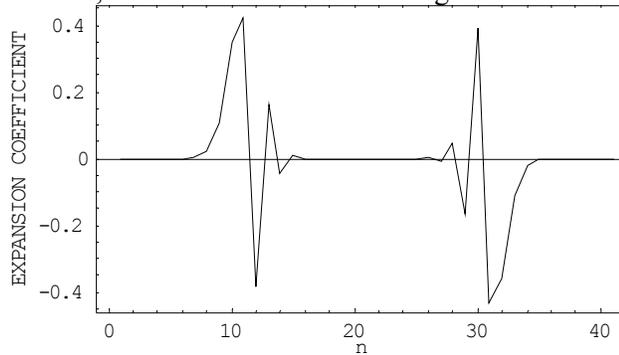

Figure 4a. The coefficients for a fragmented excited odd symmetry state for 40 particles in the double well for $\alpha$ = 4.

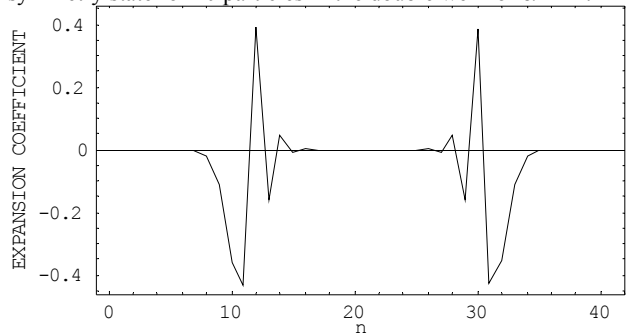

*Figure 4b*. The even symmetry pair of the state of Fig. 4a.

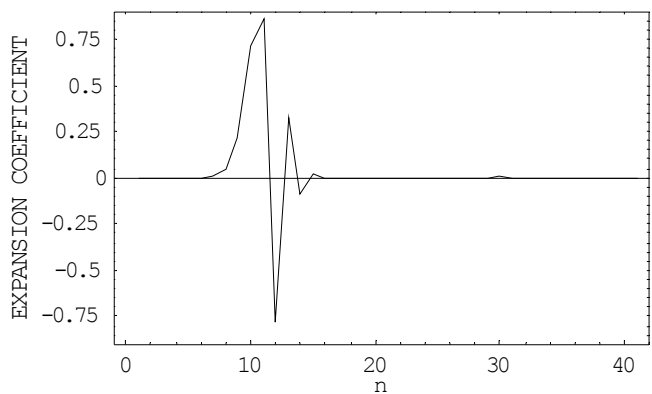

Figure 4c. A localized linear combination of the even and odd states similar to those of Figs. 4a,b, but at an even higher energy. A sign change would give localization in the opposite well.

## 3.2 Natural Orbitals and Occupation Numbers

The wavefunctions illustrated in the coefficient space of Figures 3 and 4, indicate that the wave functions for the ground and excited states of the double well problem are of two fundamentally different types depending on whether they lie below or above the crossover ridge on the energy correlation diagram of Figure 2. We now look at the occupation numbers for a series of differing values of $\alpha$. These were determined by construction and diagonalization of the 1-matix using the numerically determined coefficients of the eigenfunctions of the Hamiltonian for N= 20 particles, as in the energy correlation diagram (Figure 2), for $\alpha$ running from 2 to 5. Symmetry of the double well potential requires that in the AO basis all natural orbitals are of the form $1/(\sqrt{2})(1, \pm 1)$, namely even or odd in the symmetric co-ordinate, x. Linear combinations of degenerate states, for example as in Figure 4c, above the crossover ridge give MOs of form (1,0) or (0,1) and are fully localized. Thus, there are no surprises in the NOs as they are symmetry determined MOs. Or, should one now say COs, *condensate orbitals*? We resist this temptation.

However, the corresponding occupation numbers are quite revealing. In Figures 5a,b, for $\alpha=2$; 6a,b, for $\alpha=3$; 7a,b, $\alpha=4$ and, 8a,b, $\alpha=5$; the occupation numbers for all of the eigenfunctions of $H$ are shown. The "a" figures show the occupation numbers, $\eta_j$, for the even (1,1) NOs, and the "b" figures show the $\eta_j$ for the odd, (1,-1), NOs. In all cases the abscissa is simply the number of the energy level with the ground state labeled as j, running from 1 to 21, the energies increasing monotonically with the index.

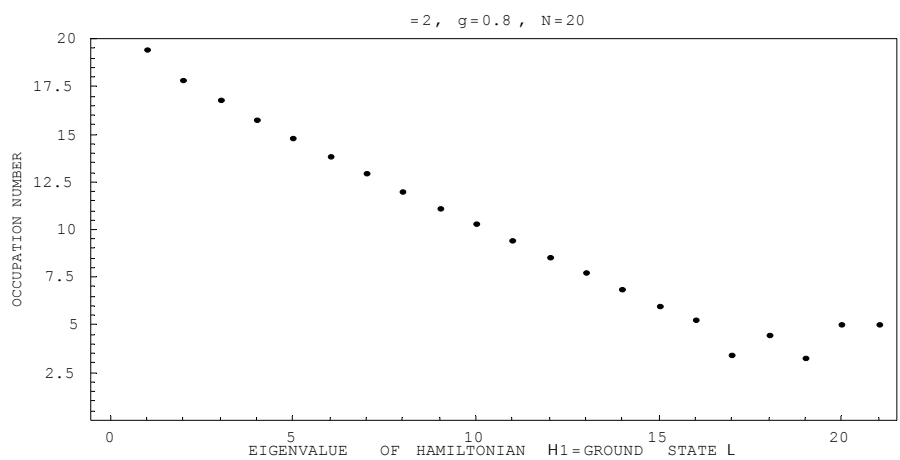

*Figure 5a.* Occupation numbers, $\eta_j$, for the $\alpha = 2$ even NOs. These begin at $\eta_j$ near 20 for the ground state, j =1 and decrease monotonically until the corresponding energies enter and pass through the crossover region near j = 17.

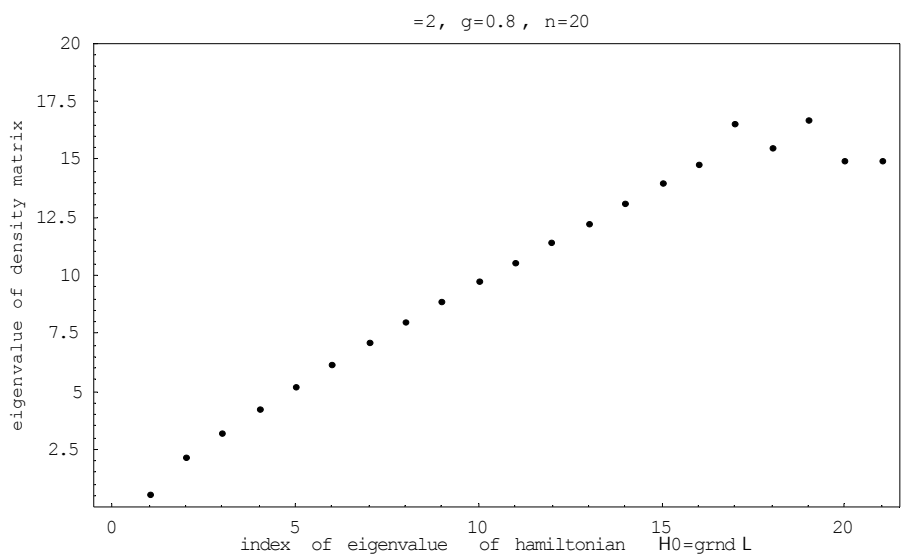

*Figure 5b.* Occupation numbers, $\eta_j$, for the $\alpha = 2$ odd NOs. These begin near 0 for the ground state, j =1 and increase monotonically until the corresponding energies enter and pass through the crossover region near j = 17.

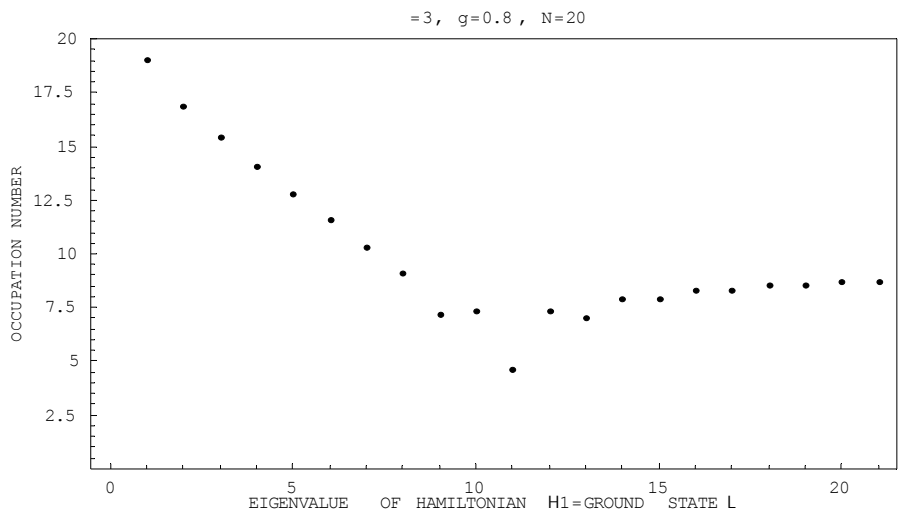

*Figure 6a.* Occupation numbers, $\eta_j$, for the $\alpha = 3$ even NOs. These begin at $\eta_j$ near 20 for the ground state, $j = 1$ and decrease monotonically until the corresponding energies enter and pass through the cross over region near $j = 10$.

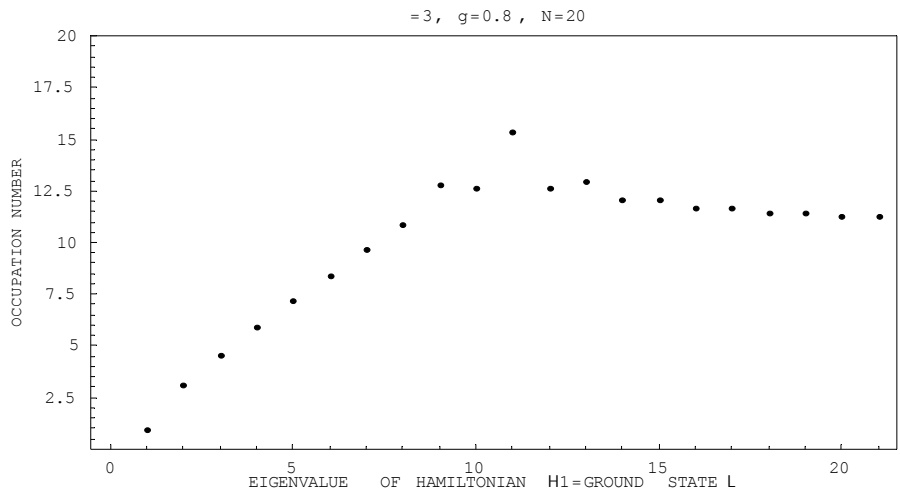

*Figure 6b.* Occupation numbers, $\eta_i$, for the $\alpha = 3$ odd NOs. These begin near 0 for the ground state, $j = 1$ and increase monotonically until the corresponding energies enter and pass through the cross over region near $j = 10$.

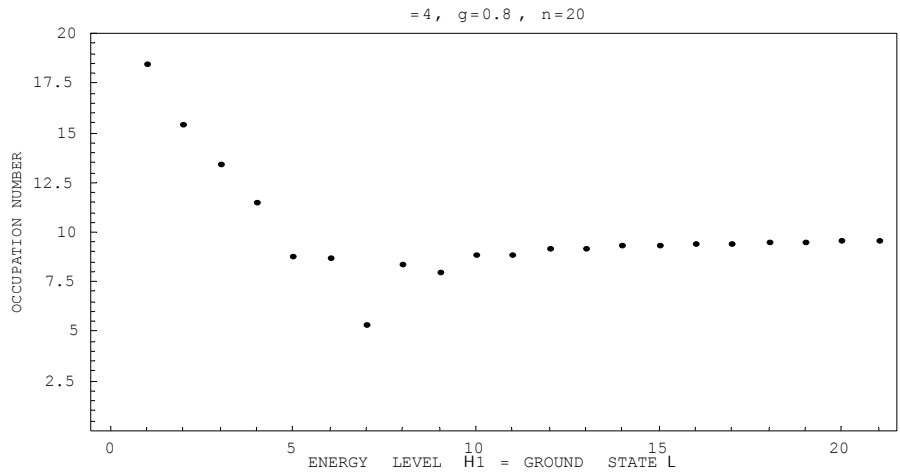

*Figure 7a.* Occupation numbers, $\eta_j$, for the $\alpha = 4$ even NOs. These begin at $\eta_j$ near 19 for the ground state, j =1 and decrease monotonically until the corresponding energies enter and pass through the cross over region near j = 6.

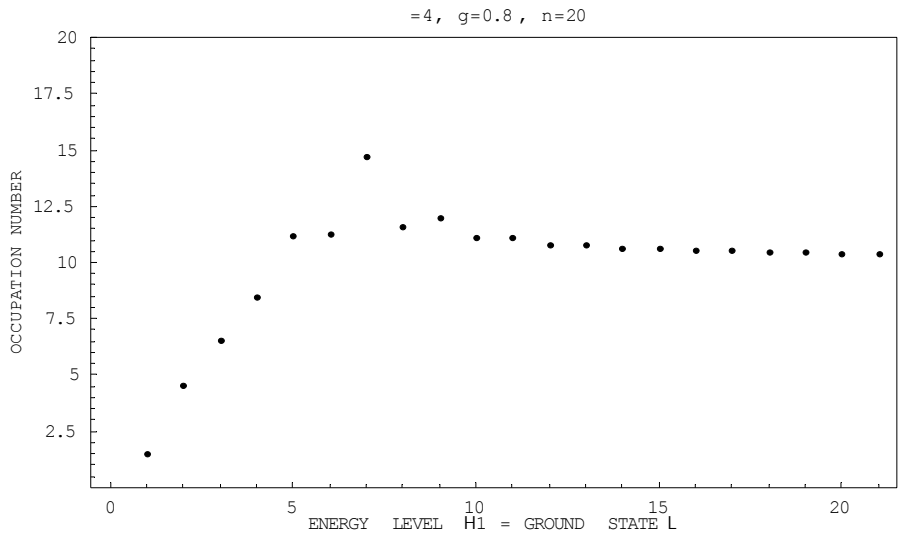

*Figure 7b.* Occupation numbers, $\eta_j$, for the $\alpha = 4$ odd NOs. These begin near 1 for the ground state, j =1 and increase monotonically until the corresponding energies enter and pass through the cross over region near j = 6.

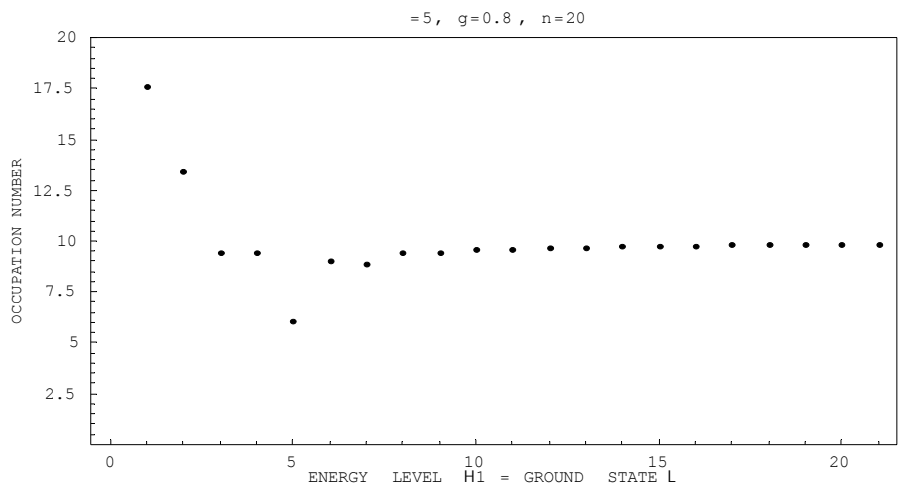

*Figure 8a.* Occupation numbers, $\eta_j$, for the $\alpha = 5$ even NOs. These begin at $\eta_j$ near 18 for the ground state, $j = 1$ and decrease monotonically until the corresponding energies enter and pass through the cross over region near $j = 4$. It is evident that all of the states with $j > 15$ have occupations numbers near 10. In this regime localized AOs provide a better zero order MOs.

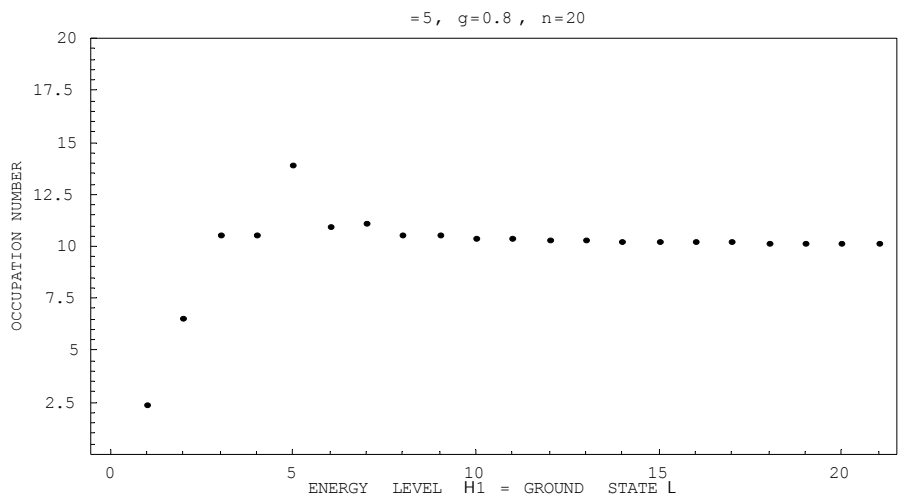

*Figure 8b.* Occupation numbers, $\eta_j$, for the $\alpha = 5$ odd NOs. These begin near 2 for the ground state, $j = 1$ and increase monotonically until the corresponding energies enter and pass through the cross over region near $j = 4$.

For all of the values of α considered, the behavior of the occupation numbers, $\eta_j$, is strikingly similar as j runs from 1 to 21, from the lowest to highest energy eigenstate of the full CI. The orbital occupancy for the ground state is close to the expected MO configuration $(\phi_R + \phi_L)^N$. The successively higher states differ from this ground configuration by transfer of atoms from $(\phi_R + \phi_L)$ to $(\phi_R - \phi_L)$. The zero order description of these states are the properly symmetrized versions of $(\phi_R + \phi_L)^{N-n} (\phi_R - \phi_L)^n$, "n" now being the number of electrons transferred from the even to the odd MO. This is precisely what a primitive MO picture would suggest, at least for energies up to the crossover region. Past the crossover energy, odd and even superpositions (which are Schrödinger Cat states) of the localized configurations $(\phi_R)^{N-n}(\phi_L)^n$ are more appropriate. These localized zero order states are *fragmented* in that fixed numbers of atoms are on a given side of the double well. As tunneling is further decreased, as a function of increasing α, the crossover region rapidly approaches the ground state, indicating that as the barrier separating the wells further increases that even the ground state will fragment, with the resulting ground state being $(\phi_R)^{N/2} (\phi_L)^{N/2}$. This is in complete agreement with the Spekkens-Sipe analysis of ground state fragmentation. Note that, in this limit, the case of even and odd N must be considered separately. We only discuss N even, herein.

For intermediate values of α, the NO analysis clearly measures the macroscopic occupancy of the ground state, and the eventually also macroscopic transfer of electrons to the higher energy odd MO, and past the crossover region the transition to localized orbitals. The crossover region is thus the boundary for a localization-delocalization transition. Interestingly, there is another interpretation of this transition, one which requires introduction of a new phase dependent fourier transformed version, $\Psi(\theta)$, of the occupation number wave function $|\Psi\rangle$ of Eqn. (3.1). This new representation will also make clear the origin of the coherent state number-phase uncertainty relationship $\Delta N \Delta\theta \geq 2p$.

## 4. Transformation to a Phase Representation: The Tunneling Problem as a Physical Pendulum

A phase representation of the wavefunction is introduced in Section 4.1 via a fourier transform. This purely mathematical step is made *physical* in Section 4.2 where a phase difference in the single particle condensate wave function on opposite sides of the double well is seen to drive a current, leading to Josephson-type oscillations of particles back and forth between the two wells. This suggests pendulum-like oscillatory motion. Both pictures lead to phase-number uncertainty relations. Both such relations are actually the same, although confusion can arise as the language describing them is, at first reading, very different. In Section 4.3 an old trick, the

Husimi transform, is used to project classical phase space information from individual quantum stationary states, and the pendulum interpretations of Section 4.2 are seen to appear automatically, even though there is no explicit time dependence anywhere in the discussion.

## 4.1 A Transformed Representation of the Wavefunction

In ordinary quantum theory a familiar transformation is between the co-ordinate and momentum representations. A harmonic oscillator wave function in co-ordinate space, Q, $\Psi_n(Q)$, may be fourier transformed to give the momentum space counterpart

$$\Psi_n(P) = \frac{1}{2\pi} \int_{-\infty}^{+\infty} e^{-iPQ} \Psi_n(Q) dQ \ . \tag{4.1}$$

Interestingly, as with proper scaling, the harmonic oscillator Hamiltonian is essentially $H_{osc} = P^2/2 + Q^2/2$, the P and Q space wavefunctions are, not surprisingly given the symmetric role of P and Q in the Hamiltonian, mathematically identical functions, each being a Gaussian times a Hermite polynomial in the appropriate P or Q variables. An equivalent transform of the occupation number representation for the fully correlated double well BEC, may be defined. The discrete vector of coefficients $\{c^j_n\}$, which we refer to as the n-representation may be used to define a θ-representation wave function, via a discrete analog of the fourier transform of Eqn (4.1):

$$\Psi_j(\theta) \equiv \sum_{n=0}^{N} c_n^j e^{-in\theta} \ . \tag{4.2}$$

This is a projection from the coefficient space vectors labeled by the discrete index "n" via the fourier functions exp(-$i$nθ). Figure 9a-c shows the θ-representation wave functions corresponding to the n-representation functions show in Figure 3. From the earlier remark about the P and Q oscillator functions being the same, we should not be overly surprised. The fourier transform of an oscillator function is an oscillator function! But perhaps the reader *is* surprised: what is oscillating in the double well BEC?

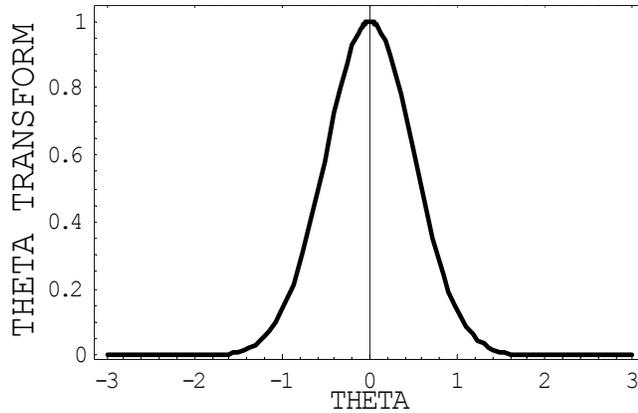

*Figure 9a.* Ψ(θ) for the α = 4 ground state of the double well BEC, plotted as a function of θ. This corresponds to the coefficient list of Figure 3a. The θ-transform of that Gaussian, is, of course, a Gaussian.

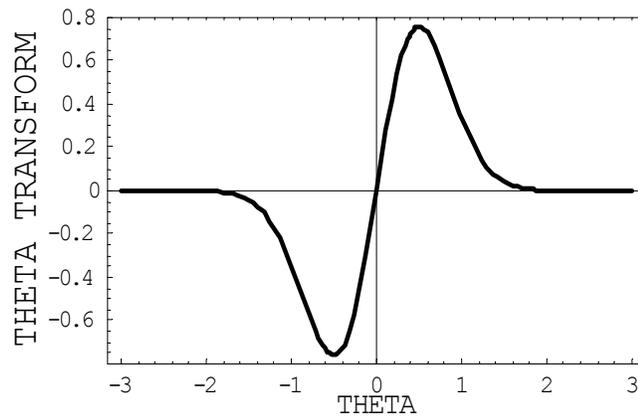

Figure 9b. Ψ(θ) for the α = 4 first excited state of the double well BEC, plotted as a function of θ. This corresponds to the coefficient list in of Figure 3b. The Q-transform of the oscillator function is still an oscillator function.

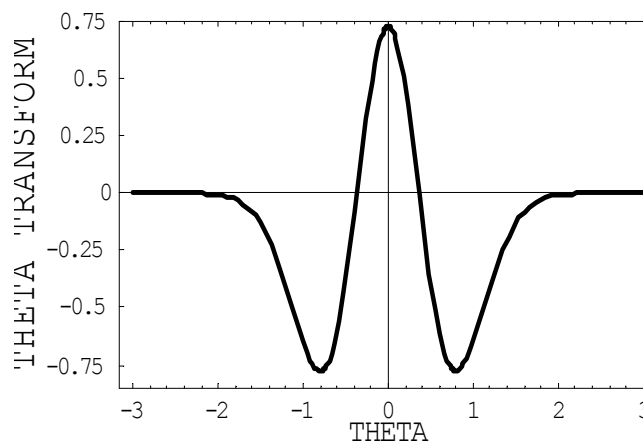

*Figure 9c.* Ψ(θ) for the α = 4 second excited state of the double well BEC, plotted as a function of θ. This corresponds to the coefficient list of Figure 3c.

Now looking on the opposite side of the crossover ridge, Figure 10 shows the θ-representation of the localized n-representation function of Figure 4c. Just as the localized n-representation function was not an oscillator like function, its θ-transfrom is perhaps even more unexpected. Note at once that we might well expect a fourier type "uncertainty" $\Delta n\Delta\theta \geq 2\pi$ from the definition of Eqn (4.2). Localization in "n", in one of the wells or another, will then correspond to delocalization in θ. Figure 10 clearly indicates this, as $\Psi(\theta)$ is now nearly constant over the full range of angle coordinate, indicating that all angles are almost equally probable, were we able to measure the angle probability distribution $P(\theta) \sim |\Psi(\theta)|^2$.

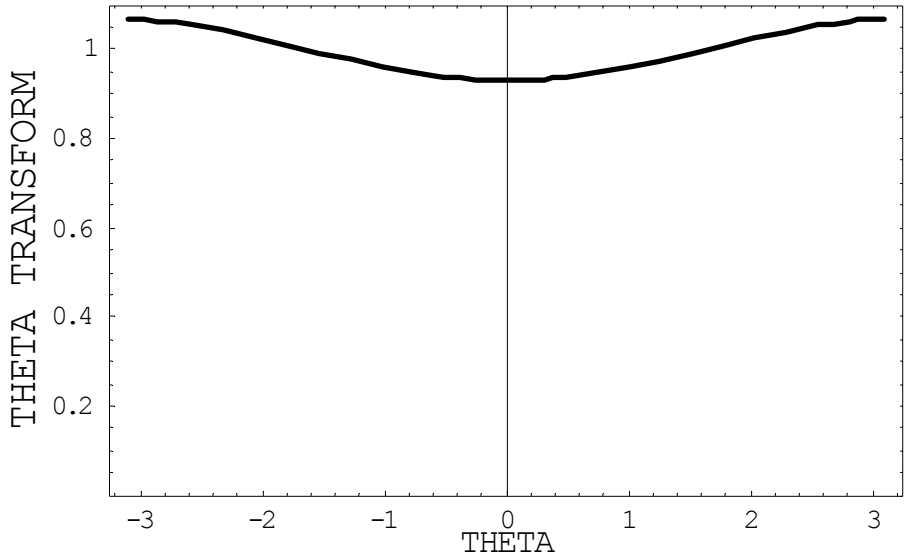

*Figure 10.* $|\Psi(\theta)|$ for the $\alpha = 4$ localized state shown in Fig. 4c, plotted as a function of θ. In this case neither the $\{c_n\}$ nor θ-representations of the wavefunction are oscillator functions. As clearly illustrated in this θ-plot, the narrow n-distribution of Fig. 4c results in a broad, almost flat $\Psi(\theta)$: this is number-phase uncertainty in action.

Summarizing this fourier (n,θ) picture: 1) we have two equivalent representations of the same quantum state, $|\Psi>$: $\Psi(\theta)=<\theta|\Psi>$, and $c_n = <n|\Psi>$. 2) These phase and number representations are related by the fourier transform of Eqn. (4.2). 3) A phase-number uncertainty arises in the usual manner of fourier uncertainties: squeeze one variable and the other broadens…as in Figure 10 where $\Delta n$ is small and $\Delta\theta$ large. The minimal uncertainty state is, as always, a Gaussian in both variables, and corresponds to the oscillator ground state.

## 4.2 The Josephson Effect and the Physical Pendulum

The transformation between the n- and θ-representations is discussed at length by Anderson and in monographs, Anderson (1984, 1966, 1964),

Tinkham (1996), and Tilley and Tilley (1990) to cite only a few examples. Is this more than a simple change of representations, as in the familiar coordinate and momentum representations of ordinary quantum mechanics, which have identical content?

Namely, may we impose, or discover, the physical nature of n and $\theta$ and understand their intertwined relationship? This is not a novel question, and it is addressed in the above three Anderson references, and originates in Anderson's elucidation of the Josephson effect. Josephson (1962) envisaged a situation where two (homogeneous) superconductors were separated by a "tunneling junction" (now referred to as a Josephson junction). Were the separate pieces of condensate to have identical densities, their condensate wavefunctions, see Eqn (2.20), might still have a relative phase offset $\theta$, so that $|N-n,n\rangle$ differs in phase $\exp(i(n-m)\theta)$ from $|N-m,m\rangle$, as a relative phase $\exp(i\theta)$ is introduced for each particle moved between wells, with the sign depending on the direction of such transfer. Josephson then suggested (Josephson, 1962) that a current would flow through the junction, and have a magnitude depending on a tunneling matrix element, $E_0$, and the phase offset $\theta$:

$$\text{Josephson Current} \quad E_0 \sin(\theta). \quad (4.3)$$

Such a current, see Anderson (1964, 1966), Tinkham (1996), Tilley and Tilley (1990) for fuller discussion, corresponds to a potential energy

$$V(\theta) = E_0(1 - \cos(\theta)). \quad (4.4)$$

Noting that the current causes a flow of particles through the junction, Anderson identified this with $n = N_R - N_L$, the difference between the number of particles on the Left and Right of the junction, and suggested that the dynamics might be well described by the classical Hamiltonian:

$$H(n,q) = \boldsymbol{K}n^2 + V(\theta). \quad (4.5)$$

If $V(\theta)$ is that of Eqn (4.5), then a natural interpretation is that the difference variable n is the analog of a momentum, and thus "$\boldsymbol{K}n^2$" is a kinetic energy, $\boldsymbol{K}$ being a constant, and the Hamiltonian of Eqn (4.5) is then that for a physical pendulum with $(n,\theta)$ being the conjugate generalized variables. The physical pendulum is a point mass held at a fixed distance from a pivot by a rigid rod. In a gravitational field two types of motion are possible: small oscillations about the potential minimum, where a Taylor expansion of $\cos(\theta)$ gives

$$V(\theta) = E_0 \theta^2/2, \quad (4.6)$$

and we see that the potential is harmonic, $E_0$ plays the role of a spring constant, and we expect small amplitude harmonic oscillations about $\theta = 0$. On the other hand, if the kinetic energy $\boldsymbol{K}n^2$ is greater than $E_0$, see Eqn (4.4), the classical motion will not be small oscillations about $\theta = 0$, but hindered and then free rotation as $\boldsymbol{K}n^2$ continues to increase. For the Josephson effect the constant $\boldsymbol{K}$ is $e^2/(2C)$, see, Anderson (1964), Tinkham (1996), as charge moves across the barrier, with a restoring propensity depending inversely on the junction capacitance, C. For the Spekkens-Sipe Hamiltonian in our parametrization, $\boldsymbol{K}$ gT$_0$, and $E_0$ gT$_1$, to within additive constants, and for small n. This identification immediately makes clear that small tunneling (i.e. small T$_1$) lowers the energy barrier for the pendulum, allowing early onset of free rotation as a function of increasing total energy. Such free rotation well describes the system at energies higher than the crossover energies of Fig. 2. Large $\alpha$, weak tunneling, then corresponds to a low energy of the crossover, and the apparent double degeneracy of the energy levels is essentially that of the double degeneracy of a particle on a circular ring. Quantization of the physical pendulum of Eqn (4.5) then follows from the correspondence rule

$$n \emptyset - i \ / \ \theta, \qquad (4.7)$$

making clear the quantum mechanical origin of the $(\Delta n \Delta \theta)$ uncertainty.

Summarizing the physical content of this Josephson-Anderson interpretation, within the context of the double well BEC: 1) A phase offset of $\theta$ for a single particle moving from one well to the other will cause a Josephson current to flow….as the condensate consists of repulsive particles, this current will ultimately reverse, and particles will flow back and forth in the frictionless quantum system. This is the pendulum undergoing small oscillations, and as there is a time dependent number of particles on either side of the potential barrier, we have a particle uncertainty. 2) Phase-number uncertainty for a stationary state of the fixed N double well system now takes on a specific physical meaning. Exact knowledge of n, the difference in particle numbers in the two wells implies a fragmented system, and then we have no knowledge of the phase relationship of the wavefunctions across the barrier. Conversely, should we wish to have maximal knowledge of the phase difference between the two wells, the spread in n will be maximal. Minimal (n,$\theta$) uncertainty arises in the ground state, where both the n and $\theta$ wave functions are Gaussians, and then it is possible that $\Delta n \Delta \theta = 2p$. This is all well illustrated in the comparisons of Figures 3,4 with 9,10.

# 4.3 Phase-Number Coherent States and the Husimi Projection into Classical Phase space

Schrödinger (1926), reprinted in English translation (1982), in one of the founding papers of wave mechanics noted that quantum time evolution of a displaced harmonic oscillator ground state led to a minimal uncertainty and non-spreading state, whose center followed a classical phase space (q,p) trajectory. Such a state is peculiar to harmonic time evolution, and is nowadays referred to as a coherent state. Such coherent states may be used to project, following Husimi (1940), see also, Gutzwiller (1990), Martens (1989), classical phase space (q,p) behavior from stationary state quantum wavefunctions in either the q or p representations. They also form a starting point for discussions of quantum optics, where the conjugate variables are then N and θ, the number of photons and the phase the electromagnetic wave. This N,θ coherent state representation of the electromagnetic field was introduced by Glauber, see Loudon (1973). Anderson, as in Sections 2.4 and 4.2, introduced it in his early discussions of ODLRO and the Josephson effect. Following Perleman (1986) the (q,p) coherent state is defined as

$$|\alpha\rangle = e^{-|\alpha|^2/2} \sum_{n=0}^{\infty} \frac{\alpha^n}{\sqrt{n!}} |n\rangle, \qquad (4.8)$$

where |n> is the n$^{th}$ eigenstate of the scaled variable harmonic oscillator Hamiltonian H(q,p) = p$^2$/2 + q$^2$/2, and α = q + *i*p, is a complex number, not to be confused with the α of the double well Hamiltonian, whose real and imaginary parts are the phase space variables (q,p). For our purposes, a more useful representation of the state |q+*i*p> is via its co-ordinate (q') or momentum (p') representations.

$$\langle q' | \alpha = q + ip \rangle = \pi^{-1/4} e^{i\sqrt{2}pq'} e^{-(q'-\sqrt{2}q)^2/2} \qquad (4.9)$$

$$\langle p' | \alpha = q + ip \rangle = \pi^{-1/4} e^{-i\sqrt{2}qp'} e^{-(p'-\sqrt{2}p)^2/2} \qquad (4.10)$$

These are interpreted as being Harmonic oscillator ground states displaced to the point (q,p) in the classical phase space. Following Schrödinger, their time (harmonic) evolution preserves the shape of the probability distribution, and its center follows the classical phase space orbit starting at the displaced point (q,p). These representations allow projection of classical phase space information from p or q representation wavefunctions: thus for example, as in Husimi (1940)

$$\langle q+ip \mid \Psi \rangle = \int_{-\infty}^{+\infty} \Psi^*(q')\langle q' \mid p+iq \rangle dq', \quad (4.11)$$

or

$$\langle q+ip \mid \Psi \rangle = \int_{-\infty}^{+\infty} \Psi^*(p')\langle p' \mid p+iq \rangle dp'. \quad (4.12)$$

These projections are the Husimi transforms of semi-classical quantum mechanics, and may be profitably used in the interpretation of wavefunctions in terms of their underlying phase space classical dynamics. Equations (4.11) or (4.12) may be used to construct phase space probability distributions, P(q,p) ª | <q+ip|Ψ> |². The Husimi transform is also a Gaussian averaged Wigner-Weyl transform of the wavefunction, a point of view not developed herein (but see, for example, Gutzwiller, 1990). Note that Steel and Collett (1998) use *Bloch sphere* Wigner-Weyl projections in their analysis of the double well problem. What we do here is rather simpler.

The (n,θ) analogs of Eqns. 4.9, 4.10 are

$$\langle \theta' \mid \alpha = \theta + in \rangle = \pi^{-1/4} e^{i\sqrt{2}n\theta'} e^{-(\theta'-\sqrt{2}\theta)^2/2} \quad (4.13)$$

$$\langle n' \mid \alpha = \theta + in \rangle = \pi^{-1/4} e^{-i\sqrt{2}\theta n'} e^{-(n'-\sqrt{2}n)^2/2} \quad (4.14)$$

The Husimi transform of an n-representation vector $c^j_n$ is thus:

$$\langle \theta + in \mid \Psi_j \rangle = \sum_{n'} c^j_{n'} e^{-i\sqrt{2}\theta n'} e^{-(n'-\sqrt{2}n)^2/2} \quad (4.15)$$

where, following Anderson, n' = $N_R$ − $N_L$ , rather than being the simpler Left-Right particle counter of Eqn. (2.14). This is simply a shift in origin. These forms assume that the oscillator Hamiltonian is now H(n,θ) = $n^2/2$+ $\theta^2/2$, namely that appropriately scaled variables are used throughout. Figure 11 shows the Husimi probability projection

$$P_j(n,\theta) = \left| \langle \theta + in \mid \Psi_j \rangle \right|^2 \quad (4.16)$$

into the classical (n,θ) phase space of two of the number representation eigenstates of the Spekkens-Sipe Hamiltonian, one state being below the crossover ridge, and one above.

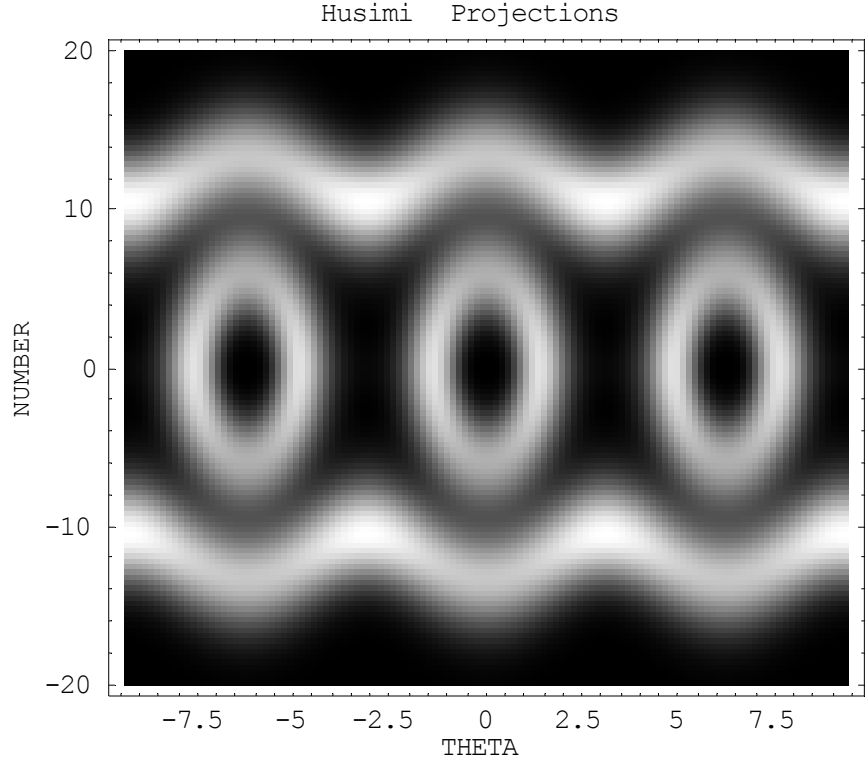

*Figure 11.* Husimi projection into the classical (n,θ) phase space of stationary state wavefunctions from above and below the crossover ridge.

These Husimi probability distributions completely confirm the Anderson physical pendulum analysis: below the classical separatrix, dividing oscillatory motion from hindred rotation, we have a periodic elliptical classical phase space orbit; above the crossover the orbits are no longer periodic, leading, rather, to ever increasing values of θ. Note also that 20 to 40 total particles (in a two basis function Hilbert space) are all that are needed to reach this semi-classical limit, wherein the underlying classical dynamics is evident. Figure 12 shows the classical orbits run at the energies of the eigenstates corresponding to the states shown in Figure 11.

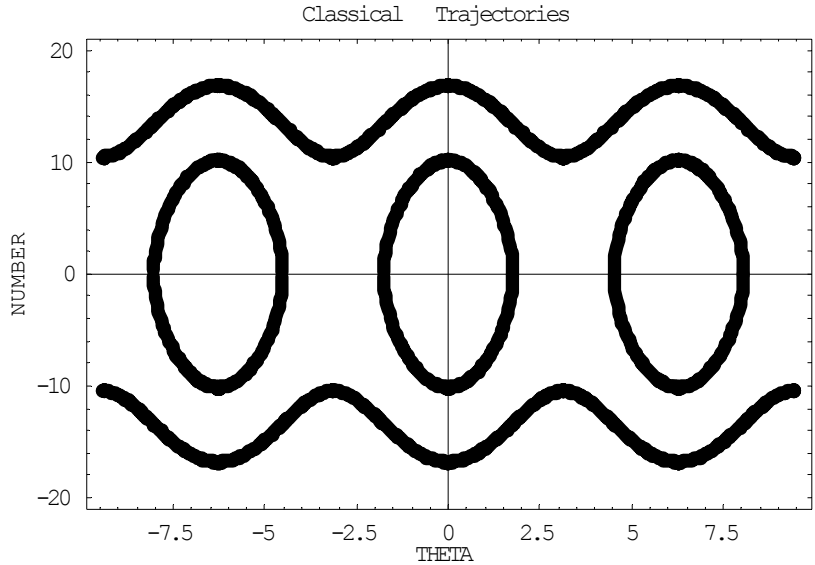

*Figure 12*. One periodic and oscillatory orbit (replicated three times, as the system is periodic in θ), and two hindered rotational orbits for a physical pendulum, with the classical Hamiltonian of Eqn (4.5). In this simple case these classical phase space orbits are simply constant energy contours. It is evident, from the comparison of Figs. 11 and 12, that the quantum stationary states for the double well BEC contain an underlying semi-classical phase space dynamics identical to that of the classical physical pendulum.

The only difference between Figures 11 and 12 is that three classical phase space orbits are shown in Fig. 12, while Fig. 11 contains the Husimi transforms of two stationary state wavefunctions. One closed orbit trajectory is shown for the periodic motion, and then a pair with identical energies, but opposite angular evolutions. These latter two correspond to (hindered) classical rotation in the two opposite directions; quantum mechanically these are the fragmented states. Were the Husimi transform of a localized state, such as that of Figure 4c, taken, the analog of one of the single hindered rotating classical trajectories would be obtained, as in Figure 10. In that case the clearly narrow distribution in n is complemented by the essentially complete uncertainty in θ. It will be interesting to make a similar two degree of freedom Husimi probability projection, as first carried out by Martens (1989) for a two degree of freedom molecular system, for a three coupled well BEC where, not unexpectedly, Salmond et. al. (2001) and for a 1 and ½ degree of freedom physical pendulum system, and Franzosi et. al., (2002a,b) for a full 2 degree of freedom system, have found evidence for chaotic dynamics. Such would be the expected generic situation for one driven or two coupled non-linear pendula, see for example, Gutzwiller (1996), Blümel and Reinhardt (1997).

# 5. Macroscopic Quantum Tunneling, a Natural Orbital-Occupation Number Analysis

Recently, there have been several reports of quantum superposition of macroscopically occupied quantum states. In all cases, rather than in gaseous BECs, as discussed here, these have been observed in superconducting current loops, with Josephson junctions inserted. van der Wal et. al., (2000); Freidman et. al. (2000); and Yu et. al. (2002) have observed avoided crossings of θ-space double well quantum states. These systems correspond to three wells of the type discussed above coupled in a circle. In these Josephson systems the effective potential V(θ) has two non-equivalent wells which may be tuned into a symmetric double well via through-the-superconducting-loop magnetic fields. The result is that quantum states initially localized in the two wells become degenerate and mix. What is actually observed in the experiments is the single photon absorption spectrum yielding the splitting between the even and odd superpositions of macroscopic quantum states as a function of the double well asymmetry. In this case the Bosons are Cooper pairs, and it is suggested that superpositions of macroscopic quantum states with millions of such Cooper pairs have been observed. What does this mean? What we see, by analysis of analogous tunneling splttings in our double well model, is that very small number of such Bosons have changed their single particle quantum state, and an NO analysis shows that only a small number of particles are involved in the spectra observed. The observed tunneling splittings are thus the few-body excitation spectrum: seemingly paradoxically the macroscopic quantum superposition of large numbers of particles is actually controlled by a very few of these.

Although the double well BEC discussed here is not precisely equivalent to the Josephson systems mentioned above, the analogous macroscopic superposition states may be indeed observed in the excited states of the double well model discussed above. To find such strong mixing of near degenerate states, one only need look at states at energies just above the cross-over ridge of Fig. 2. Too far above this ridge, such tunneling splittings will be too small of easily observe, but by tuning the tunneling, via α, and choosing the quantum states carefully, double well states with small, but not too small, splittings may be easily found, and their behavior observed by adjusting the one of the two well depth parameters $\varepsilon_{LL}$ or $\varepsilon_{RR}$, which we have previously taken to be equal. We have chosen to take $\varepsilon_{LL}(\tau) = \varepsilon_{LL} + 0.1\tau$, and $\varepsilon_{RR}(\tau) = \varepsilon_{LL} - 0.1\tau$, and to let $\tau$ vary from -1/2 to 1/2, $\tau = 0$ then being the point of the minimal avoided crossing between the two levels. Figure 13 shows some of the lower lying levels for $\alpha = 8$ plotted as a function of $\tau$ for 50 particles. The first noticeable avoided crossing is for the splitting between levels 4 and 5. A magnification of the avoided crossing for these same two levels is shown in Figure 14. A natural-orbital and occupation number analysis of the states of Fig. 14 allows us to ask and answer the question: in a transition between these levels how many particles actually change quantum state? For single photon spectroscopy to successfully observe such a splitting one would guess that the answer must be a fairly small number, or selection rules would make it unlikely that the transition would be observed. In principle the transition from an even to an odd macroscopic state might only involve the transition of a single particle from an even to an odd orbital, or the reverse.

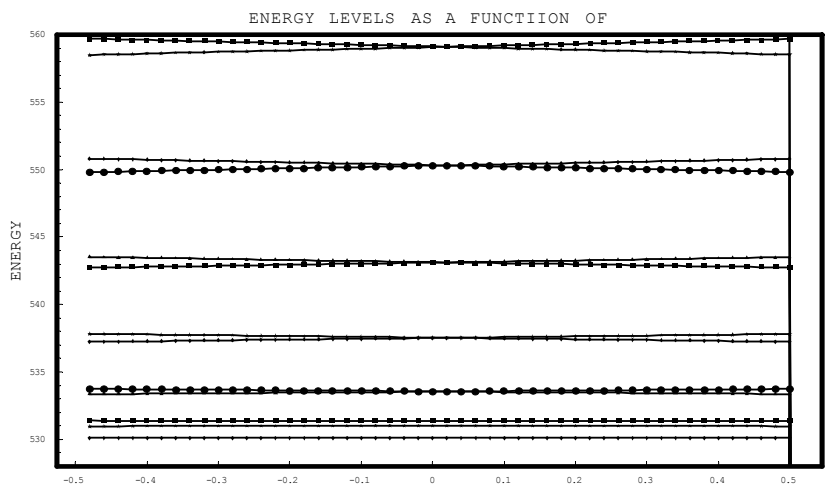

*Figure 13.* Energy levels for the asymmetric double well, $\varepsilon_{RR} = \varepsilon_{LL} + 0.1\tau$, for $\tau$ running from $-1/2$ to $+1/2$. Levels are for 50 particles with a high barrier, $\alpha = 8$, implying that levels 4,5 are already above the crossover ridge, and are near degenerate at $\tau = 0$. A magnification of this lowest energy avoided crossing is shown in the following figure.

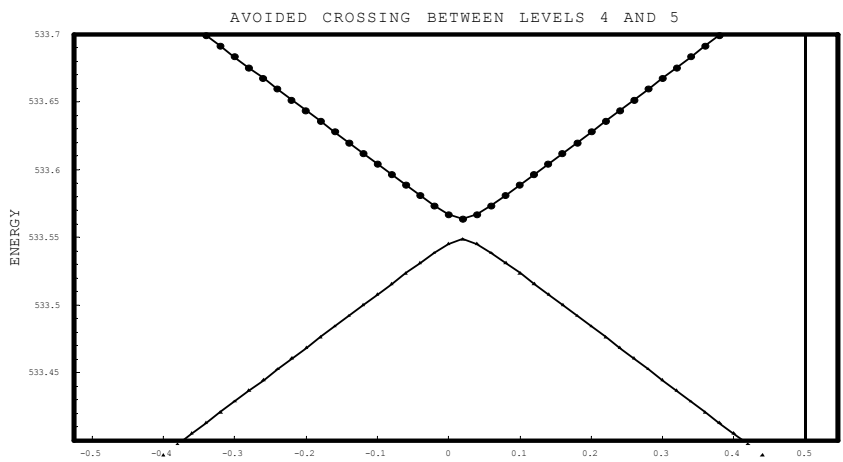

*Figure 14.* Avoided crossing of levels 4,5 as a function of well asymmetry parameter, $\tau$, for N = 50, $\alpha = 8$. Both states are hindered rotor states, above the crossover ridge. A natural orbital/occupation number analysis follows.

Thus even though large numbers of particles might be involved in superconducting current loops, the experimentally observed splittings might only involve the transition of a single, or a relatively small number of Bosons. What, then, are the occupation numbers for the states involved in such an avoided crossing as shown in Figure 14? These, and similar occupation numbers for (approximately) equivalent avoided crossings, for varying numbers of particles are shown in Table I.

Table I. Occupation numbers for nearly degenerate states for 26, 50, 100, 200, 400, and 800 particles. In each case $\tau = 0$, $\alpha = 8$, and the occupation numbers are shown for the even (1,1) and odd (1,-1) MOs as labeled by the pair of energy levels involved in the avoided crossing. Although the details do vary over the states shown, it is evident that the occupation numbers generally change by amounts small compared to the total number of particles. For example in the N = 400 case, only "about 3" particles actually change from even to odd MOs in the transition between the states 18 and 19. Thus although macroscopic quantum superpositions are indeed involved, the spectra may involve only few body transitions. As $\alpha$ increases, the number of particles involved in the "split" macroscopic superposition pair approaches 1, and eventually 0 as the condensate is fully fragmented.

| Number of particles | State numbers | occupation # (1,1) | occupation # (1,-1) |
|---|---|---|---|
| 26 | 2 | 13.5185 | 12.4815 |
|  | 3 | 10.6135 | 15.3865 |
| 50 | 4 | 23.8434 | 26.1566 |
|  | 5 | 23.0011 | 26.9989 |
| 100 | 6 | 45.2315 | 54.7685 |
|  | 7 | 43.6142 | 56.3858 |
| 200 | 10 | 84.5568 | 115.443 |
|  | 11 | 81.9966 | 118.003 |
| 400 | 18 | 157.629 | 242.371 |
|  | 19 | 154.557 | 245.443 |
| 800 | 32 | 277.862 | 522.138 |
|  | 33 | 250.013 | 549.987 |

The occupation numbers in Table I. show that each natural orbital is certainly macroscopically occupied, that is both have a major fraction of all particles. The occupation numbers of successive levels in the avoided crossings region give an indication that at the curve crossing, $\tau = 0$, the state mixing may well involve only a few particles changing from even to odd MOs. In fact, as the tunneling barrier is further raised, the number of particles moving from an even to an odd (or vice versa) MO converges zero, representing a fully fragmented state with zero tunneling splitting. This is consistent with the fact that observed spectra in the papers, van der Wal et.

al. (2000), Freidman et. al. (2000), and Yu et. al. (2002), are single photon spectra. One photon will not couple states wherein too many particles move co-operatively. Thus, for example in Freidman et. al. (2000) where the macroscopically mixed states are oppositely propagating supercurrents, it is likely that single photon transitions, and angular momentum conservation, imply that what is being observed is the transition of one or a small number of particles from one macroscopically occupied configuration to another. An occupation number analysis will give further insight into that more complex situation. But, at last, the "single particle excitation spectrum" of a quantum fluid has been identified. This interpretation differs from that of the above experimentalists, who are used to thinking in terms of collective modes of superfluids and superconductors. What we seem to have here, in contrast, are the types of single particle excitations characteristic of those many parts of atomic, molecular, and nuclear physics.

## 6. Summary and Conclusions

The natural orbitals and occupation numbers, introduced by Löwdin in 1955 are found to provide a most useful diagnostic for the states of a gaseous Bose condensate in a model double well potential. To describe the fragmentation, or dissociation, of such a system correlations are needed, as the Hartree-GP mean field picture does not provide a correct description. Once correlations are introduced the Löwdin occupation numbers provide a clear diagnostic as to what states are occupied by the condensed Bosons, and in doing so make it evident that the transitions observed in recent macroscopic tunneling experiments may be, in fact, part of the few-body, rather than the more usual collective excitations of such systems. Similar techniques should be applicable to looking at fragmentation and number squeezing in multi-well gaseous BECs and in superconducting current loops. This is of particular interest as such systems are possible realizations of the qubits of quantum computers.

## 7. Acknowledgements

The authors acknowledge helpful conversations with J. Brand and K. Mahmud, and the support of the U.S. National Science Foundation through Grant PHY- 0140091. The paper is dedicated to the memory of P.-O. Löwdin whose enthusiastic and seminal work on natural-orbitals, and the related N-representability problem, inspired, trained, and stimulated many generations of theoretical scientists. It is a pleasure to be able to apply these ideas to illuminate the physics of a novel set of systems.